\documentclass[a4paper,12pt]{article}
\usepackage{amsmath,amssymb}
\usepackage{epic}
\newcommand{\wh}{\widehat}

\newcommand{\wt}{\widetilde}

\begin{document}
\title{A new two-dimensional lattice model\\ that is
``consistent around a cube''}

\author{Jarmo Hietarinta\\Department of Physics, University of
  Turku\\FIN-20014 Turku,
  FINLAND\\ e-mail:~Jarmo.Hietarinta@utu.fi}

\maketitle

\begin{abstract}
  For two-dimensional lattice equations one definition of
  integrability is that the model can be naturally and consistently
  extended to three dimensions, i.e., that it is ``consistent around a
  cube'' (CAC). As a consequence of CAC one can construct a Lax
  pair for the model. Recently Adler, Bobenko and Suris conducted a
  search based on this principle and certain additional assumptions.
  One of those assumptions was the ``tetrahedron property'', which is
  satisfied by most known equations. We present here one lattice
  equation that satisfies the consistency condition but does not have
  the tetrahedron property. Its Lax pair is also presented and some
  basic properties discussed.
\end{abstract}

\section{Introduction}
Within the field of integrable dynamics there have recently been
interesting developments in the study of integrable {\em difference}
equations (an overview of the topic can be obtained from the
proceedings mentioned in \cite{history}).  These include both the
discussion on what is a proper definition of integrability, or whether
the different suggested definitions actually agree or not.  Numerous
integrable (in some sense) difference equations have been proposed by
discretizing known ODE's and PDE's in a way that retains some good
properties. Once a definition of integrability has been proposed one
can also try to search for all equations having the chosen property.

\begin{figure}
\begin{center}
\setlength{\unitlength}{0.0004in}
{\renewcommand{\dashlinestretch}{30}
\begin{picture}(3482,3813)(0,-10)

\put(1275,2708){\circle*{150}}
\put(625,2808){\makebox(0,0)[lb]{$x_{[2]}$}}
\put(-2700,2708){\makebox(0,0)[lb]{$x_{n,m+1}\equiv x_{[2]}$}}

\put(3075,2708){\circle*{150}}
\put(3375,2808){\makebox(0,0)[lb]{$x_{[12]}$}}
\put(4800,2808){\makebox(0,0)[lb]{$x_{n+1,m+1}\equiv x_{[12]}$}}

\put(1275,908){\circle*{150}}
\put(825,1008){\makebox(0,0)[lb]{$x$}}
\put(-2000,1008){\makebox(0,0)[lb]{$x_{n,m}\equiv x$}}

\put(3075,908){\circle*{150}}
\put(3300,1008){\makebox(0,0)[lb]{$x_{[1]}$}}
\put(5000,1008){\makebox(0,0)[lb]{$x_{n+1,m}\equiv x_{[1]}$}}

\drawline(275,2708)(4075,2708)
\drawline(3075,3633)(3075,0)
\drawline(275,908)(4075,908)
\drawline(1275,3633)(1275,0)
\end{picture}
}
\end{center}
\caption{The lattice map is defined on a elementary square of the
  lattice}\label{F:1}
\end{figure}
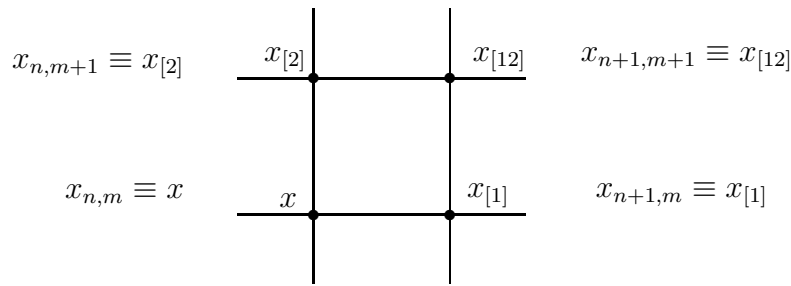

In this paper we consider integrable difference equations defined on a
2-dimensional lattice.  We assume that the  lattice is rectangular and
infinite, and concentrate on some square in it (see Figure \ref{F:1})
(We normally use subscripts in square brackets to indicate shifts in
the indices, but for composite expression we sometimes also use $\wt{\ 
}$ for shift in the 1-direction and $\widehat{\ }$ for a shift in the
2-direction.)  In general the map is given in terms of a
multi-linear equation relating the four corner values in Figure
\ref{F:1}:
\begin{eqnarray}
&&K\, x x_{[1]} x_{[2]} x_{[12]} + L_1\, x x_{[1]} x_{[2]} + L_2\, 
x x_{[1]} x_{[12]}
 + L_3\, x x_{[2]} x_{[12]} + L_4\, x_{[1]} x_{[2]} x_{[12]}\nonumber\\
&& + P_1\, x x_{[1]} + P_2\,
 x_{[1]} x_{[2]} + P_3\, x_{[2]} x_{[12]} + P_4\, x_{[12]} x +
P_5\, x x_{[2]} + P_6\, x_{[1]} x_{[12]}\nonumber\\
&& + Q_1\, x + Q_2 \,x_{[1]} + Q_3\, x_{[2]}
 + Q_4\, x_{[12]} + U\equiv 
Q_{12}(x,x_{[1]},x_{[2]},x_{[12]};p_1,p_2)=0.
\label{basicmap}
\end{eqnarray}
Here the coefficients $K,\,L_\nu,\,P_\nu,\,Q_\nu,\,U$ may depend on
the two spectral parameters $p_1,p_2$. If any 3 of the corner values
are given then the fourth one can be obtained as a rational expression
of the other three.  One can therefore propagate any staircase-like
initial value line to cover the whole plane\cite{NC}.

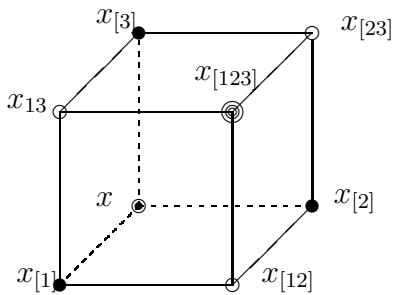
\begin{figure}[b]
\begin{center}
\setlength{\unitlength}{0.0005in}
{\renewcommand{\dashlinestretch}{30}
\begin{picture}(3482,3013)(0,-10)
\put(450,1883){\circle{150}}
\put(-100,1883){\makebox(0,0)[lb]{$x_{13}$}}

\put(1275,2708){\circle*{150}}
\put(825,2708){\makebox(0,0)[lb]{$x_{[3]}$}}

\put(3075,2708){\circle{150}}
\put(3375,2633){\makebox(0,0)[lb]{$x_{[23]}$}}

\put(2250,83){\circle{150}}
\put(2550,8){\makebox(0,0)[lb]{$x_{[12]}$}}

\put(1275,908){\circle{150}}
\put(1275,908){\circle*{90}}
\put(825,908){\makebox(0,0)[lb]{$x$}}

\put(2250,1883){\circle{150}}
\put(2250,1883){\circle{220}}
\put(2250,1883){\circle{80}}
\put(1850,2108){\makebox(0,0)[lb]{$x_{[123]}$}}

\put(450,83){\circle*{150}}
\put(0,8){\makebox(0,0)[lb]{$x_{[1]}$}}

\put(3075,908){\circle*{150}}
\put(3300,833){\makebox(0,0)[lb]{$x_{[2]}$}}

\drawline(1275,2708)(3075,2708)
\drawline(1275,2708)(450,1883)
\drawline(450,1883)(450,83)
\drawline(3075,2708)(2250,1883)
\drawline(450,1883)(2250,1883)
\drawline(3075,2633)(3075,908)
\dashline{60.000}(1275,908)(450,83)
\dashline{60.000}(1275,908)(3075,908)
\drawline(2250,1883)(2250,83)
\drawline(450,83)(2250,83)
\drawline(3075,908)(2250,83)
\dashline{60.000}(1275,2633)(1275,908)
\end{picture}
}
\end{center}
\caption{Given the values at the black circles, one should get a unique
  value for  $x_{[123]}$, even though there are three possible way to
  compute it.}\label{F:2}
\end{figure}

How should integrability be defined for such maps?  In \cite{NW} the
following definition of integrability (``Consistency Around the
Cube'', CAC) was proposed (see Figure \ref{F:2}): Adjoin a third
direction (therefore assuming $x=x_{n,m,k}$) and use the same map (but
with different spectral parameters) also in planes corresponding to
indices 1,3 and 2,3. That is, the map given in \eqref{basicmap}
contains shifts and parameters associated with directions 1,2 now the
same should be done with directions 3,1 and 2,3, furthermore, on on
the parallel shifted planes we use identical maps.
We assume that the values $x,\,x_{[1]},\,x_{[2]},\,x_{[3]}$ at
black circles in Figure \ref{F:2} are given, then the values at open
circles are uniquely determined using the relevant map, but the value
at $x_{[123]}$ can be computed in 3 different ways, and they must give
the same result.  In other words:
\begin{eqnarray*}
\text{solve $x_{[12]}$ from } &  
Q_{12}(x,x_{[1]},x_{[2]},x_{[12]};p_1,p_2)=0,\\
\text{solve $x_{[23]}$ from } &  
Q_{23}(x,x_{[2]},x_{[3]},x_{[23]};p_2,p_3)=0,\\
\text{solve $x_{[31]}$ from } &  
Q_{31}(x,x_{[3]},x_{[1]},x_{[31]};p_3,p_1)=0,
\end{eqnarray*}
then $x_{[123]}$ computed from
\begin{eqnarray*}
Q_{12}(x_{[3]},x_{[31]},x_{[23]},x_{[123]};p_1,p_2)=0,&& \text{ or }\\
Q_{23}(x_{[1]},x_{[12]},x_{[31]},x_{[123]};p_2,p_3)=0,&& \text{ or }\\
Q_{31}(x_{[2]},x_{[23]},x_{[12]},x_{[123]};p_3,p_1)=0,&&
\end{eqnarray*}
should be the same. The functions $Q_{ij}$ could be in principle be
different, but are usually assumed to be identical.

The above idea (and diagram) resembles, e.g., the 3D Bianchi diagram
that is obtained from consistency of Moutard transformations (see
\cite{NS}), but it is here used in a different context: the diagram
introduces a condition for lattice maps defined on {\em squares}
rather than for Moutard transformations defined on {\em edges}.
Furthermore, this CAC principle is a {\em constructive} definition of
integrability in the sense that it leads algorithmically to a Lax
pair.

The following maps are well known examples that have CAC
property\cite{history,NC}:
\begin{enumerate}
\item Lattice KdV: $(p_1-p_2+x_{[2]}-x_{[1]})(p_1+p_2+x-x_{[12]})
=p_1^2-p_2^2,$
\item Lattice MKdV: $p_1(x x_{[2]}-x_{[1]}x_{[12]})
=p_2(x x_{[1]}-x_{[2]}x_{[12]}),$
\item Lattice SKdV: $(x-x_{[2]})(x_{[1]}-x_{[12]})p_2^2
=(x-x_{[1]})(x_{[2]}-x_{[12]})p_1^2.$
\end{enumerate}

\section{Searching for integrable lattices}
Now that the definition of integrability has been given one may ask
for a listing of all integrable models. The complete classification of
CAC maps is in fact a formidable open problem.  The set of equations
that needs to be solved can be derived from above: comparing the
different forms for $x_{[123]}$ and collecting the various
coefficients of $x$ and its shifts, we get $2\times 375$ functional
equations for the coefficient functions (in practice we have often
used $3\times 375$ equations for a symmetric approach). The equations
are polynomial in the coefficient functions appearing in
\eqref{basicmap}, but the functions depend on different pairs of the
three spectral parameters $\lambda_i,\,i=1,2,3$ and all three
$\lambda_i$'s appear in each equation. The equation list itself takes
35MB to store.

In \cite{ABS} the consistency equations were solved in the case where
the map $Q$ is the same on all planes, and under two additional
assumptions:
\begin{itemize}
\item symmetry: 
\begin{eqnarray}&&Q(x,x_{[1]},x_{[2]},x_{[12]};p_1,p_2)=\varepsilon
  Q(x,x_{[2]},x_{[1]},x_{[12]};p_2,p_1)\nonumber\\
 &&=\sigma Q(x_{[1]},x,x_{[12]},x_{[2]};p_1,p_2),
\quad \varepsilon,\sigma=\pm 1,
\label{symm}
\end{eqnarray}
\item  ``tetrahedron property'': $x_{[123]}$ does not depend on $x$.
\end{itemize}
With these assumptions the authors were able to get a full
classification resulting with 9 models.

The tetrahedron assumption is indeed satisfied by most of the well known
models, but one can nevertheless ask whether it is a fundamental or
essential property and whether there are any nontrivial models that do
not satisfy it. It should be immediately observed that the more or
less trivial models 
\begin{equation}\label{triv}
x_{[1]}x_{[2]}-x x_{[12]}=0,\text{and }y-y_{[1]}-y_{[2]}+y_{[12]}=0,
\end{equation}
($x=e^y$) do have CAC property but {\em not} the tetrahedron property, in
fact one quickly finds that
\begin{equation}
x_{[123]}=x_{[1]}x_{[2]}x_{[3]} x^{-2},\text{ and }
y_{[123]}=y_{[1]}+y_{[2]}+y_{[3]}-2y,
\end{equation}
respectively.  One may now ask whether these models have nontrivial
extensions. 

The symmetry assumptions \eqref{symm} lead to 2 possibilities
w.r.t.~$\sigma$, the one with $\sigma=-1$ is:
\begin{eqnarray}
Q&=&a_1(p_1,p_2)(xx_{[1]}x_{[2]}-xx_{[2]}x_{[12]}-xx_{[1]}x_{[12]}
+x_{[1]}x_{[2]}x_{[12]})\nonumber \\&&+
a_2(p_1,p_2)(xx_{[12]}-x_{[1]}x_{[2]})
+a_3(p_1,p_2)(x-x_{[1]}-x_{[2]}+x_{[12]}).
\label{sigma}
\end{eqnarray}
In \cite{ABS} it was observed that in this symmetry class there are no
integrable cases with tetrahedron property, which is true.  The ansatz
\eqref{sigma} forms, 
fortunately, a rather simple class and a direct computation shows that
it contains one new integrable map
\begin{align}
  x x_{[1]} x_{[2]} - x x_{[1]} x_{[12]} -& x x_{[12]} x_{[2]}
  + x_{[1]} x_{[2]} x_{[12]}\nonumber\\
  + (x_{[1]} x_{[2]}-x x_{[12]})& ( p_1 + p_2)- 
p_1 p_2(x-x_{[1]}-x_{[2]}+ x_{[12]})=0,
\label{bowtie}
\end{align}
with two parameters. Since it can also be written as 
\[
\begin{array}{cccccccl}
 x & x_{[1]} & x_{[2]}& & & &- pq & x_{[12]}\\
 - x & x_{[1]} & x_{[12]} &\phantom{w}+
 (p+q) & x_{[1]} & x_{[2]} \phantom{w}&+ pq & x_{[2]}\\
- x & x_{[2]} & x_{[12]}&\phantom{w}-(p+q ) &x & x_{[12]} 
\phantom{w}& + pq & x_{[1]}\\
+ x_{[1]} & x_{[2]} & x_{[12]}& & & & - pq & x
\end{array}\quad =0,
\]
we call it ``the bow-tie model''.  Model \eqref{bowtie} is, however,
still M\"obius-equivalent to the first model in \eqref{triv} by $x\to
-(p_1x+p_2)/(x+1)$ [although this transformation does not trivialize
the maps on the other planes, since it depends explicitly on
$p_1,p_2$].  Therefore we decided to search for possible extensions.
This was done perturbatively, with completely general 1st and 2nd
order extensions to the bow-tie map (this work was done with REDUCE
3.7\cite{reduce}, and 1GB memory). In this approach it is not
necessary to specify the details of the parameter dependence -- indeed
the result was surprising in this respect. From the outcome of this
exercise we noticed certain weaker symmetry properties among the
coefficient functions and when the (nonperturbative) computations were
done with these properties (and translational invariance) the
following solution was found:
\begin{align}
  Q(x,x_{[1]},x_{[2]},x_{[12]};e_1,o_1;e_2,o_2)
\equiv\hskip 2cm&\nonumber\\
 x x_{[1]} x_{[2]} (o_1 - o_2) - x
    x_{[1]} x_{[12]} (e_1 - o_2) -& x  x_{[2]} x_{[12]} (o_1 - e_2)
    + x_{[1]} x_{[2]} x_{[12]}(e_1 - e_2)\nonumber\\
    + (x_{[1]} x_{[2]}-x x_{[12]})& (e_1 o_1 - e_2 o_2)\nonumber\\
    + (x x_{[2]} o_1-x_{[1]} x_{[12]}e_1) (e_2 - o_2)
    & + ( x_{[2]} x_{[12]} e_2 - x x_{[1]} o_2 )(e_1 - o_1)\nonumber\\
    - x_{[12]} e_1 e_2 (o_1 - o_2) + x_{[2]} e_2 o_1 (e_1 - o_2) & +
    x_{[1]} e_1 o_2 (o_1 - e_2)
    - x o_1 o_2 ( e_1 - e_2)\nonumber\\
    &=0.\label{eq:Hmap}
\end{align}
This new model is the main result of this paper.  It is interesting to
note that in \eqref{eq:Hmap} there are {\em two} parameters $o_i,e_i$
in each direction, and that if $e_i=o_i$ then $o_1-o_2$ factors out
leaving \eqref{bowtie}. The parameters with different indices must all
be different, otherwise the map factorizes. This model obeys the CAC
property, in fact one finds
\begin{align*}
\text{num}&(x_{[123]})=
-x (x_{[1]}+o_1) (x_{[2]}+o_2) (x_{[3]}+o_3) (o_1-o_2) (o_2-o_3) (o_3-o_1)\\&
+(x_{[1]}+o_1) (x_{[2]}+o_2) (x_{[3]}+o_3)\\&\quad\times 
\left[(e_1 e_2+e_3 o_3) o_3 (o_1-o_2)+
(e_2 e_3+e_1 o_1) o_1 (o_2-o_3)+
(e_3 e_1+e_2 o_2) o_2 (o_3-o_1)\right]\\&
+(x+e_3) (x_{[1]}+o_1) (x_{[2]}+o_2) o_3 (o_1-o_2) (e_2-o_3) (o_3-e_1)\\&
+(x+e_1) (x_{[2]}+o_2) (x_{[3]}+o_3) o_1 (o_2-o_3) (e_3-o_1) (o_1-e_2)\\&
+(x+e_2) (x_{[3]}+o_3) (x_{[1]}+o_1) o_2 (o_3-o_1) (e_1-o_2) (o_2-e_3),\\
\text{den}&(x_{[123]})=
(x_{[1]}+o_1) (x_{[2]}+o_2) (x_{[3]}+o_3) \\&\quad\times 
  \left[(e_1 e_2+e_3 o_3) (o_2-o_1)+
   (e_2 e_3+e_1 o_1) (o_3-o_2)+
   (e_3 e_1+e_2 o_2) (o_1-o_3)\right]\\&
+ (x+e_3)(x_{[1]}+o_1) (x_{[2]}+o_2) (o_1-o_2) (e_1-o_3) (e_2-o_3)\\&
+ (x+e_1)(x_{[2]}+o_2) (x_{[3]}+o_3) (o_2-o_3) (e_2-o_1) (e_3-o_1)\\&
+ (x+e_2)(x_{[3]}+o_3) (x_{[1]}+o_1) (o_3-o_1) (e_3-o_2) (e_1-o_2),
\end{align*} 
which are symmetric under permutations of $1,2,3$, and the explicit
$x$-dependence demonstrates violation of the tetrahedron property of
\cite{ABS}.

\section{Symmetries}
The model \eqref{eq:Hmap} is invariant under the M\"obius transformation
\begin{eqnarray*}
X&\to& \phantom{-}\frac{\alpha X+\beta}{\gamma X+\delta},\quad X=x,
x_i, x_{ij}, x_{ijk}\\
P&\to& -\,\frac{\alpha P-\beta}{\gamma P -\delta},\quad 
P=e_i\text{ or }o_i,
\end{eqnarray*}
It has the symmetries
\begin{eqnarray}
Q(x,x_{[1]},x_{[2]},x_{[12]};e_1,o_1,e_2,o_2)&=&
-Q(x,x_{[2]},x_{[1]},x_{[12]};e_2,o_2,e_1,o_1),
\\
Q(x,x_{[1]},x_{[2]},x_{[12]};e_1,o_1,e_2,o_2)&=&
-Q(x_{[1]},x,x_{[12]},x_{[2]};o_1,e_1,e_2,o_2).
\label{nsymm2}
\end{eqnarray}
These are similar to \eqref{symm} with $\varepsilon=\sigma=-1$, but
note that in \eqref{nsymm2} there is an exchange between the two
parameters related to direction 1. Indeed, it seems that the extension
of the parameter space from 1 to 2 dimensional (in each direction of
the lattice) allows the introduction of a nontrivial exchange symmetry,
resulting with the nontrivial model.  (Recall that if $e_i=o_i$ the
model simplifies to \eqref{bowtie}.)

The map \eqref{eq:Hmap} can also be written in the following rational form
\begin{equation}\label{ratform}
\frac{x+e_2}{x+e_1}\;\frac{x_{[12]}+o_2}{x_{[12]}+o_1}=
\frac{x_{[1]}+e_2}{x_{[1]}+o_1}\;\frac{x_{[2]}+o_2}{x_{[2]}+e_1}.
\end{equation}
(The B\"acklund equation presented in Eq.~(5.11) of \cite{NS2} is
similar, but the distribution of the terms is in fact different.)
A form equivalent to \eqref{ratform} is
\[
\frac{o_2+x_{[12]}}{o_2+x_{[2]}}\;\frac{e_2+x}{e_2+x_{[1]}}=
\frac{o_1+x_{[12]}}{o_1+x_{[1]}}\;\frac{e_1+x}{e_1+x_{[2]}},
\]
and comparing these one observes duality under
\begin{equation}
o_2\leftrightarrow x,\,e_2\leftrightarrow x_{[12]},\,
o_1\leftrightarrow x_{[1]},\,e_1\leftrightarrow x_{[2]}.
\end{equation}

We also note that \eqref{eq:Hmap} has the simple solution
\[
x_{n,m,p}=c+(e_1-o_1)n+(e_2-o_2)m+(e_3-o_3)p.
\]

\section{Lax pair}
An important property of the CAC definition of integrability is that
it provides a method of constructing a Lax pair. The recipe was
provided by Nijhoff in \cite{N2}: We consider the three maps on planes
12, 23, 31.  The third direction is taken as auxiliary (spectral) and
the system is linearized in the corresponding variable $x_{[3]}$ and
its shifts. We also use notation $o_3=\mu,\,e_3=\lambda$.

Solving for $x_{[31]}$ from
$Q(x,x_{[3]},x_{[1]},x_{[31]};\lambda,\mu,e_1,o_1)=0$ yields
\[
x_{[31]}=\frac{\begin{array}{l}x_{[3]} [
x x_{[1]} (o_1 - \mu )
 + x o_1 (\lambda  - \mu )
 + x_{[1]} (e_1 o_1 - \lambda  \mu )
 + \lambda  o_1 (e_1 - \mu )]\\
\hskip 1cm + x x_{[1]} \mu  ( - e_1 + o_1)
 + x o_1 \mu  ( - e_1 + \lambda )
 + x_{[1]} e_1 \mu  ( - \lambda  + o_1)\end{array}}
{\begin{array}{l} x_{[3]} [x_{[1]} ( - e_1 + \lambda )
 + x ( - \lambda  + o_1)
 + \lambda  ( - e_1 + o_1)]\\
\hskip 1cm + x x_{[1]} (e_1 - \mu )
 + x (e_1 o_1 - \lambda  \mu )
 + x_{[1]} e_1 (\lambda  - \mu )
 + e_1 \lambda  (o_1 - \mu )
\end{array}}
\]
and this is then linearized by introducing $f,g$ by 
\begin{equation}
x_{[3]}=\frac{f}{g},\quad x_{[23]}=\frac{f_{[2]}}{g_{[2]}},
\quad x_{[31]}=\frac{f_{[1]}}{g_{[1]}},
\end{equation}
resulting in 
\begin{align*}
f_{[1]}=&\kappa_1 \, f \left[x x_{[1]} (\mu-o_1)
 + x o_1 (\mu-\lambda)
 + x_{[1]} (\lambda\mu-e_1 o_1)
 + \lambda  o_1 (\mu-e_1)\right]\\
 + &\kappa_1 \, g\left[x x_{[1]} \mu  (e_1 - o_1)
 + x o_1 \mu  (e_1 - \lambda )
 + x_{[1]} e_1 \mu  (\lambda  - o_1)\right],\\
g_{[1]}=&\kappa_1 \,  f \left[x_{[1]} (e_1 - \lambda )
 + x (\lambda  - o_1)
 + \lambda  (e_1 - o_1)\right]\\
 + &\kappa_1 \, g\left[x x_{[1]} (\mu-e_1)
 + x (\lambda\mu-e_1 o_1)
 + x_{[1]} e_1 (\mu-\lambda)
 + e_1 \lambda  (\mu-o_1)\right].
\end{align*}
Here the overall separation factor $\kappa_i$ may contain $x,x_{[i]}$.  We
write this, and the corresponding equation obtained from 
$Q(x,x_{[2]},x_{[3]},x_{[23]};e_2,o_2,\lambda,\mu)=0$ in matrix
form,
\begin{equation}
\phi=\begin{pmatrix}f\\g\end{pmatrix},\quad
\phi_{[i]}=\begin{pmatrix}f_{[i]}\\g_{[i]}\end{pmatrix},\quad
\phi_{[i]}=L_i\phi,
\end{equation}
with
\begin{equation}
L_i=\kappa_i(x_{[i]}+o_i)(x+\lambda)
\begin{pmatrix}\mu &\mu e_i\\-1 & -e_i\end{pmatrix}
+\kappa_i(x_{[i]}+\lambda)(x+e_i)
\begin{pmatrix}-o_i &-\mu o_i\\1 & \mu\end{pmatrix}.
\end{equation}
Now from the consistency condition
$\left(\phi_{[1]}\right)_{[2]}=\left(\phi_{[2]}\right)_{[1]}$ we get
the matrix equation
\begin{equation}\label{basicLax}
\wh L_1 L_2=\wt L_2 L_1,
\end{equation}
which is satisfied modulo the map \eqref{eq:Hmap}, provided that the
separation constants $\kappa_i$ satisfy
\begin{equation}
\frac{\kappa_1\wt \kappa_2}{\wh \kappa_1\kappa_2}=
\frac{(x_{[2]}+e_1)(x+e_2)(x_{[2]}+\lambda)}
{(x+e_1)(x_{[1]}+e_2)(x_{[1]}+\lambda)}.
\end{equation}
A simple solution to this is
\begin{equation}
\kappa_i=\frac1{(x+e_i)(x+\lambda)},
\end{equation}
leading to
\begin{equation}
L_i=\dfrac{x_{[i]}+o_i}{x+e_i}
\begin{pmatrix}\mu &\mu e_i\\-1 & -e_i\end{pmatrix}
+\dfrac{x_{[i]}+\lambda }{x+\lambda}
\begin{pmatrix}-o_i &-\mu o_i\\1 & \mu\end{pmatrix}
\end{equation}
A similarity transformation with 
$S=\left(\begin{smallmatrix}1& 0\\ \mu^{-1} &1\end{smallmatrix}\right)$
simplifies this further to
\begin{equation}
L_i'=\begin{pmatrix}(\mu-e_i)\dfrac{x_{[i]}+o_i}{x+e_i} &
\mu e_i\dfrac{x_{[i]}+o_i}{x+e_i}-\mu o_i\dfrac{x_{[i]}+\lambda}{x+\lambda}\\
0&(\mu-o_i)\dfrac{x_{[i]}+\lambda}{x+\lambda}\end{pmatrix}.
\end{equation}

\section{Summary}
The ``consistency around the cube'' definition of integrability is
very transparent since it directly leads to the Lax pair. It also
defines a clear search problem whose complete solution is,
unfortunately, still beyond our computational abilities.  The
integrable model \eqref{eq:Hmap} or \eqref{ratform} presented in this
paper is the first result in our search project.  This model has
several interesting features:
\begin{itemize}
\item  It does {\em not} satisfy  the tetrahedron property.
\item It has {\em two} parameters in each direction.
\item It is dual under interchange of variables and parameters.
\end{itemize}
Further properties of this model will be discussed elsewhere.

In this general class of lattice models (multi-linear, consistent
around a cube) there are probably still many other models to be found.

\section*{Acknowledgements}
The author would like to thank Frank Nijhoff for numerous discussions.

\end{document}